\documentclass[twocolumn,showpacs,showkeys,preprintnumbers]{revtex4-1}
\usepackage{amssymb}

\usepackage{amsfonts}

\usepackage{latexsym}

\usepackage{dcolumn}

\usepackage{amsmath}

\usepackage{graphicx}

\usepackage{psfrag,pstricks}

\newcommand{\eqn}{\ref}

\begin{document}

\title{Generation of motional nonlinear coherent states and their superpositions via intensity-dependent coupling of a cavity field to
a micromechanical membrane }

\author{ Sh. Barzanjeh$^{1}$ }
\email{shabirbarzanjeh@yahoo.com}

\author{M. H. Naderi$^{2}$}
\email{mhnaderi@phys.ui.ac.ir}
\author{ M. Soltanolkotabi$^{2}$}
\email{soltan@sci.ui.ac.ir}
\affiliation{ $^{1}$ Department of Physics, Faculty of Science, University of Isfahan, Hezar Jerib, 81746-73441, Isfahan, Iran\\
$^{2}$Quantum Optics Group, Department of Physics, Faculty of Science, University of Isfahan, Hezar Jerib, 81746-73441, Isfahan, Iran}

\date{\today}

\begin{abstract}

We propose a theoretical scheme to show the possibility of generating motional nonlinear
coherent states and their superposition  for an undamped vibrating micromechanical
membrane inside an optical cavity. The scheme is based on an intensity-dependent
coupling of the membrane to the radiation pressure field. We show that if the cavity field
is initially prepared in a Fock state, the motional state of the membrane may evolve from
vacuum state to a special type of nonlinear coherent states. By examining the nonclassical
properties of the generated state of the membrane, including the quadrature squeezing
and the sub-Poissonian statistics, we find that by varying the Lamb-Dicke parameter and
the membrane's reflectivity one can effectively control those properties. In addition, the
scheme offers the possibility of generating various types of the so-called nonlinear
multicomponent Schr\"{o}dinger cat sates of the membrane. We  also examine the effect of the damping
of the cavity field on the motional state of the membrane.
\end{abstract}

\maketitle

\section{Introduction}\label{Introduction}
The coupling of mechanical motion of mesoscopic or macroscopic mirrors to the electromagnetic degrees of freedom via radiation pressure\cite{cave,corbitt1} is a promising approach for applications ranging from the detection of weak forces and small displacements to fundamental studies of the transition between the quantum and the classical world \cite{kipp}. The canonical optomechanical system consists of an optical cavity where one of the end mirrors is free to move \cite{fabre}. The radiation pressure acting on the movable mirror realizes an optomechanical coupling between the incident optical modes and the various vibrational modes of the mirror. The use of optomechanical coupling has been proposed many years ago for the implementation of quantum limited measurements of mechanical forces\cite{Brag}, as the interferometric detection of gravitational waves \cite{abra} or atomic force microscopy\cite{rugar}. Since then, optomechanical systems have generated much experimental and theoretical interest\cite{zhang, kle,gigan,bhatt,cohadon}. They offer the prospect of realizing quantum effects at a macroscopic scale \cite{marshall}, of supplying quantum sensors for applications ranging from single molecule detection \cite{yang} to gravitational wave interferometry \cite{corbitt2,courty}, for the quantum control of atomic, molecular, and optical systems \cite{treu}, and for possible new quantum information processing devices\cite{mancini1}.
The optomechanical cooling and trapping of mirrors has also recently become the subject of an intense research effort\cite{kle,gigan,bhatt,corbitt2,genes}as it offers a viable means of exploring quantum effects such as superposition and entanglement at a macroscopic scale\cite{marshall,mancini2}.

The main technical barrier to reaching the regime in which either the mechanical device itself or its readout demonstrate quantum behavior, has been difficulty of integrating ultrasensitive micromechanical devices with high-finesse optical cavities. Recently, it has been developed a novel type of optomechanical system, the so-called membrane-in-the middle geometry, for addressing this issue in which the mechanical degree of freedom is a flexible, partially transparent dielectric membrane placed inside a Fabry-Perot cavity with fixed end mirrors\cite{thompson,jay1,jay}. This has the advantage of not having to combine the flexibility needed for the mechanical oscillator with the rigidity of a high-finesse cavity mirror. Although the membrane is nearly transparent, it couples to the optical cavity dispersively. This coupling is strong enough to laser-cool a 50-nm-thick dielectric membrane from room temperature(294 K) down to 7mK\cite{thompson}. In addition, the dispersive nature of the optomechanical coupling allows for realization of a sensitive displacement squared readout of the membrane\cite{thompson}. Such a readout is a crucial requirement for measuring quantum jumps in a mechanical oscillator. It has been shown theoretically\cite{helmer} that an optomechanical system with the membrane-in-the-middle geometry can be used to detect phonon Fock states. Furthermore, for that system it has been predicted \cite{nunn} two-phonon cooling of the mechanical oscillator, squeezing of the mechanical oscillator, and squeezing of the optical output field. The extension of cavity optomechanics to multimembrane systems has also been considered\cite{bhatta}.

In connection with quantum state engineering, optomechanical systems have also attracted considerable attention because of the great possibilities they have to produce nonclassical states of both, the mechanical oscillator and cavity field\cite{cohadon,marshall,zhang,mancini2,bose,fabre,mancini3,pace,jahne}.
Similar to the case of a light field where the imprint of the quantum regime is signaled by generation of nonclassical states, the generation of nonclassical states of a mechanical oscillator can be a hallmark for quantum control of a macroscopic massive object\cite{walls}. The possibilities to generate nonclassical states, and in particular a superposition of coherent states of the quantized field are because of the Kerr-like Hamiltonian that may be obtained in the optomechanical systems\cite{bose,mancini3}. Besides, the fact that the motion of the mechanical oscillator is quantized allows the possibility to generate a large variety of nonclassical states of the mirror\cite{bose}. It is shown that the mirror can also be prepared in a Schr\"{o}dinger cat state with many components by a quadrature measurement of the cavity field after its interaction with the moving mirror. Moreover, the possibility of generating even and odd coherent states of a moving mirror has been considered\cite{zheng}. A detailed analysis of the effects of finite temperature on creating and verifying a macroscopic quantum superposition in a micro-optomechanical system has been presented in \cite{kleckner}. It has been shown that an unambiguous demonstration of a quantum superposition requires the mechanical resonator to be in or near the ground state. In Ref.\cite{zhang} the authors propose a scheme for transferring quantum states from the propagating light fields to macroscopic, collective vibrational degree of freedom of a massive mirror by exploiting radiation pressure effects. Some of other more recent proposals include the generation of squeezed states of a vibrating membrane or a movable mirror via reservoir engineering in an optomechanical setup\cite{jahne}, the creation of EPR beams in an optomechanical system\cite{zhang-qi}, and the creation of robust entanglement against increasing temperature between the optical intracavity mode and the mechanical mode of the mirror\cite{ferr}.

In recent years, there has been paid much attention to the study of a family of generalized coherent states, the so-called nonlinear coherent states(NLCSs)\cite{dematos} because of their relevance in nonlinear quantum optics. They correspond to nonlinear algebras rather than Lie algebras\cite{manko}. For a detailed discussion of the mathematical properties of NLCSs we refer the reader to Ref.\cite{manko,aniello}. The NLCSs, and their superposition exhibit nonclassical features like amplitude squeezing, sub-Poissonian statistics and self-splitting accompanied by pronounced quantum interference effects\cite{sivakumar}. These states are not mere mathematical objects. In the context of ion trap, the microscopic NLCSs may appear as stationary states of the center-of-mass motion of a trapped and laser-driven ion far from the Lamb-Dicke regime\cite{dematos}. The method makes use of the strong \textit{nonlinearities} inherent in the Jaynes-Cummings Hamiltonian for the laser-assisted vibronic interaction\cite{vogel}. Very recently, it has been introduced\cite{mahdifar} a physical scheme that allows one to prepare and control a special kind of these states, the so-called sphere-coherent states, as the motional dark states of a properly laser-driven trapped-atom system. Furthermore, it has been explored\cite{naderi} the possibility of generating various families of NLCSs in a lossless micromaser within the framework of intensity-dependent Jaynes-Cummings model.

Motivated by the above-mentioned studies on the generation of various types of nonclassical states in an optomechanical structure, in the present paper
we deal with the question of how  the NLCSs can be generated in an optomechanical system. The system under consideration is an optomechanical cavity with membrane-in-the middle geometry that consists of a high finesse cavity with two perfectly reflecting fixed end mirrors, and a partially reflective movable
middle mirror(such as a dielectric membrane)\cite{thompson,jay1,jay,borkje}. In this type of optomechanical structure, the radiation pressure and cavity detuning are periodic in the membrane displacement. This periodicity leads to an intensity-dependent interaction of the radiation pressure field with the movable membrane. As we shall see, in the absence of any damping
mechanisms, this system can lead to the generation of motional nonlinear coherent states
and their superposition for the membrane. We also study the nonclassical properties of
generated state and show that how the system parameters (in particular, membrane's
reflectivity and the Lamb-Dicke parameter) can strongly affect the dynamics of the
system under consideration. If those states are to be observed, then very good isolation of
the system from the environment is necessary. Although, by taking into account the
relevant experimental parameters \cite{thompson,jay1}, just the same considered in this paper, one can
neglect the membrane's motional damping, but one cannot neglect the cavity damping.
Therefore, in order to make the model under consideration more realistic, we also
consider the photon leakage from the cavity as a relevant source of decoherence and
examine its influence on the motional state of the membrane.

The paper is organized as follows. In Sec. II we first derive an intensity-dependent
Hamiltonian describing the coupling of the micromechanical membrane to the radiation
pressure field through  \textit{j}-phonon excitations of the vibrational sideband. Then, by
restricting our treatment to the case of  \textit{j=1} and by using the Feynman disentanglement
theorem we derive the time evolution operator of the system. In Sec. III we consider the
generation of NLCS of the membrane and investigate the nonclassical properties of the
generated state. In Sec. IV the preparation of the nonlinear multicomponent Schr\"{o}dinger
cat states of the membrane is discussed. In Sec. V we take into account the cavity field damping and study its effect on motional state of the membrane. Finally, we summarize our conclusions in Sec. VI.

\section{The nonlinear coupling of radiation pressure field to a mechanical degree of freedom}
\label{model}
\subsection{Model Hamiltonian}
As is shown in Fig.1, we consider a high finesse cavity which is detuned by the motion of a partially reflective membrane placed between two macroscopic, rigid, perfectly reflecting fixed end mirrors\cite{thompson,jay1,jay,borkje}. Unlike the standard optomechanical structures, in this type of optomechanical system the coupling between the middle membrane and the optical cavity strongly depends on the position of the membrane. This position dependence results in a cavity detuning, which is a periodic function of the membrane displacement~$x$, i.e., $\omega_c(x)=(c/L)cos^{-1}[|r_c|cos(4\pi x/\lambda)]$ where $L$ and $r_c$ are the cavity length and the field reflectivity of the membrane, respectively. The Hamiltonian of the system is given by (excluding damping and driving terms)\cite{thompson,jay1}

\begin{eqnarray}\label{1}
 \hat{H}=\hbar \omega_c(\hat x)\hat a^{\dag}\hat a +\hbar \omega_m \hat b^{\dag}\hat b,
\end{eqnarray}
where $\hat a$ and $\hat b$ are, respectively, the lowering operators for the optical and mechanical modes satisfying the commutation relations $[\hat a,\hat a^{\dag}]=1, [\hat b,\hat b^{\dag}]=1$, and $\omega_m$ is the oscillation frequency of the middle membrane. The fact that the motion of the membrane is quantized allows writing the operator of membrane's displacement $\hat x$ in terms of the Lamb-Dicke parameter $\eta=\frac{\omega_p}{L\omega_m}\sqrt{\frac{\hbar}{2m\omega_m}}$\cite{rae1,rae2} ($\omega_p$ and $m$ denote, respectively, the frequency of the incident field and the motional mass of the membrane)as
\begin{eqnarray}\label{2}
 \hat{x}=\eta \frac{2L\omega_m}{c}(b^{\dag}+\hat b).
\end{eqnarray}
 By expanding $\omega_c(\hat x)$ in term of $\hat b$ and $\hat b^{\dagger}$ we obtain
\begin{widetext}
\begin{equation}\label{3}
\begin{array}{rcl}
\omega_c(\hat x)=\frac{\pi c}{2L}-\frac{c}{2L}
\sum\limits_{m = 1}^{\infty} {\sum\limits_{k = 0}^{\frac{{m - 1}}{2}} {\frac{{\left| {r_c } \right|^m }}{m}} } \left( {\begin{array}{*{20}c}
   m  \\
   k  \\
\end{array}} \right)\frac{{(m - 1)!}}{{4^{m - 1} [(\frac{{m - 1}}{2})!]^2 }}\{ e^{i\eta \theta (m - 2k)(\hat b+\hat b^{\dagger})} + h.c.\},
\end{array}
\end{equation}
\end{widetext}
where $m=2l+1$ ($l$ is an integer number) and $\theta=2L\omega_m/c$. By using the Baker-Campbell-Hausdorff theorem in Eq.(\eqn{3}) we may rewrite the Hamiltonian of equation(\eqn{1}) in the form
\begin{eqnarray}\label{4}
 \hat{H}=\hat H_0+\hat H^j_{int},
\end{eqnarray}
where
\begin{eqnarray}\label{5}
\hat H_0=\hbar \omega_0\hat a^{\dag}\hat a +\hbar \omega_m \hat b^{\dag}\hat b,
\end{eqnarray}\\
with $\omega_0=\pi c/(2L)$ as the natural frequency of the cavity without middle membrane, describes the free Hamiltonian of the quantized cavity field and the free motion of the mechanical degree of freedom, and
\begin{eqnarray}\label{6}
\hat H_{int}^j=\hbar \hat a^{\dag}\hat a[\chi^{*}_jf_j(\hat n_b)\hat b^{j}+\chi_j(\hat b^{\dagger})^{j}f_j(\hat n_b)]\nonumber,\,(j=0,1,2,..)\\
&&
\end{eqnarray}
where
\begin{equation}\label{9}
\begin{array}{rcl}
\chi_j=\frac{c}{2L}(i\frac{4\pi }{\lambda}\sqrt{\frac{\hbar}{2m\omega_m}})^j,
\end{array}
\end{equation}
and the nonlinearity function $f_j(\hat n_b)$ is defined by(appendix A)
\begin{widetext}
\begin{equation}\label{7}
\begin{array}{rcl}
f_j(\hat n_b)=
\sum\limits_{m = 1} {\sum\limits_{k = 0}^{\frac{{m - 1}}{2}} {\frac{{\left| {r_c } \right|^m }}{m}} } \left( {\begin{array}{*{20}c}
   m  \\
   k  \\
\end{array}} \right)\frac{{(m - 1)!}}{{4^{m - 1} [(\frac{{m - 1}}{2})!]^2 }}\{ e^{-\frac{1}{2}(\eta\theta)^{2} (m - 2k)^{2}}\}\frac{\hat{n}_b!(m-2k)^{j}}{(\hat{n}_b+j)!}L^{j}_{\hat{n}_b}[(\eta\theta)^{2} (m - 2k)^{2}],
\end{array}
\end{equation}
\end{widetext}
with $\hat n_b=\hat b^{\dagger}\hat b$ and $L^{j}_{\hat{n}}$ as the associated Laguerre polynomial, describes a nonlinear coupling of the radiation pressure field with the movable membrane through \textit{j}-phonon excitations of the vibrational sideband.

\subsection{TIME EVOLUTION}
The time evolution operator corresponding to the Hamiltonian (\eqn{4}) can be evaluated easily. For this purpose let us consider the first excitation of the vibrational sideband by choosing $j=1$ in the Hamiltonian (\eqn{4}):
\begin{equation}\label{8}
\begin{array}{rcl}
\hat H=\hbar \omega_0\hat a^{\dag}\hat a +\hbar \omega_m \hat b^{\dag}\hat b+\hbar \,\hat a^{\dag}\hat a[\chi^\ast
f(\hat n_b) \hat b+\chi\hat b^{\dagger}f(\hat n_b)],
\end{array}
\end{equation}
where we have defined
\begin{eqnarray}
\chi\equiv \chi_1&=&i\frac{2\pi c}{L\lambda}\sqrt{\frac{\hbar}{2m\omega_m}},\nonumber\\
f(\hat n_b)&\equiv &f_1(\hat n_b).
\end{eqnarray}
The nonlinearity
function $f(\hat n_b)$ plays an important role in our treatment since it determines the form of nonlinearity of the intensity-dependent of the coupling between the cavity field and the membrane. We point out that in the limit of very small values of the Lamb-Dicke parameter $\eta$ and for certain values of the membrane reflectivity $r_c$ the nonlinearity function $f(n_b)$ reduces to unity. Fig.2(a) shows $f(n_b)$ as a function of $n_b$ for $\eta=0.8$ and $r_c=0.99$, while Fig.2(b) displays the behavior of $f(n_b)$ versus $n_b$ for $\eta=10^{-5}$ and $r_c=0.9$. Obviously, for $f(n_b)=1$ the Hamiltonian (\eqn{8}) reduces to the Hamiltonian of the standard opto-mechanical system\cite{bose}. Therefore, the inherent nonlinearity of the model under consideration can be attributed to the parameters of $\eta$ and $r_c$.

In the following, we consider the unitary time evolution operator in the interaction picture, corresponding to the Hamiltonian (\eqn{8}) that
reads
\begin{equation}\label{10}
\begin{array}{rcl}
\hat {U}_{int}^{(j=1)}(t)=\hat T exp\big[-\frac{i}{\hbar} \int\limits_{0}^{t} H_{int}^{(j=1)}(t')dt'\big],
\end{array}
\end{equation}
where $\hat T$ indicates the time ordering operator. By substituting Hamiltonian (\eqn{8}) in the interaction picture
 \begin{equation}\label{10.1}
\begin{array}{rcl}
 \hat H_{int}^{(j=1)}=\hbar \hat a^{\dagger}\hat a \Big(\chi^{*} f(\hat n_b)\hat b e^{-i \omega_m t}+\chi\hat b^{\dagger}f(\hat n_b)e^{i \omega_m t}\Big),
 \end{array}
\end{equation}
into Eq.(\eqn{10}) and using the Feynman operator calculus\cite{fey} to disentangling the time-ordered evolution operator(appendix B) we obtain
\begin{equation}\label{11}
\begin{array}{rcl}
\hat {U}^{(j=1)}_{int}(\tau)= e^{\beta \mu(\tau)\hat a^{\dagger}\hat a \hat B^{\dagger}}e^{-\beta^{*} \mu^{*}(\tau)\hat a^{\dagger}\hat a \hat B}\times\\
\,\,\,\,e^{i |\beta|^2\lambda(\tau)(\hat a^{\dagger}\hat a)^2 g(\hat n_b)}e^{O(\beta ^3,\hat n_b)+...},
\end{array}
\end{equation}
where by definition
\begin{equation}\label{12}
\begin{array}{rcl}
\tau=\omega_m t;\,\,\beta= \chi/\omega_m;\,\, \mu(\tau)= 1-e^{i \tau};\,\,\lambda(\tau)= t+i\mu^{*}(\tau),
\end{array}
\end{equation}
and
\begin{eqnarray}\label{13}
\hat B&=& f(\hat n_b)\hat b,\\\nonumber
g(\hat n_b)&=&(\hat n_b+1)f^2(\hat n_b)-\hat n_bf^2(\hat n_b-1).
\end{eqnarray}
Now we make use of an approximation to simplify the unitary operator in Eq.(\eqn{11}). In this approximation we keep terms up to second order in $\beta$. The accuracy of our approximations may be verified by considering the realistic values of $\beta$ based on the experimental values of the relevant parameters\cite{thompson,jay1}. For $\omega_m/2\pi=10^5 Hz,\,\,m= 50 pg,\,\,\lambda=532 nm$ and $L=0.67 cm$, we have $|\chi|\sim10 kHz$ which leads to $|\beta|\sim 0.01$. Within this approximation the time evolution operator in Eq.(\eqn{11}) reduces to
\begin{equation}\label{14}
\begin{array}{rcl}
\hat {U}^{(j=1)}_{int}(\tau)\simeq e^{\beta \mu(\tau)\hat a^{\dagger}\hat a \hat B^{\dagger}}e^{-\beta^* \mu^{*}(\tau)\hat a^{\dagger}\hat a \hat B}e^{i |\beta|^2\lambda(\tau)(\hat a^{\dagger}\hat a)^2 g(\hat n_b)}.
\end{array}
\end{equation}

Now, we assume that the membrane is initially prepared in a coherent superposition of the phononic Fock states  $|\psi(0)\rangle_m=\sum\limits_{k}C_k|k\rangle_m$ and the cavity field is initially in a coherent superposition of the photonic Fock states $|\psi(0)\rangle_f=\sum\limits_{n}D_n|n\rangle_f$. Thus the initial state vector of the system can be written as

\begin{equation}\label{15}
\begin{array}{rcl}
|\psi(0)\rangle=\sum\limits_{k,n}C_k D_n|k\rangle_m\otimes|n\rangle_f.
\end{array}
\end{equation}
The time dependent state of the system is obtained by
\begin{equation}\label{16}
\begin{array}{rcl}
|\psi(\tau)\rangle=\hat {U}^{(j=1)}_{int}(\tau)|\psi(0)\rangle.
\end{array}
\end{equation}
By substituting (\eqn{14}) and (\eqn{15}) into Eq.(\eqn{16}) we obtain
\begin{equation}\label{17}
\begin{array}{rcl}
|\psi(\tau)\rangle=\sum\limits_{n,k}C_k D_n e^{in^2\Theta g(k)} e^{\Lambda_n \hat B^{\dagger}} e^{-\Lambda_n^{*}\hat B}|k\rangle_m\otimes|n\rangle_f,
\end{array}
\end{equation}
where $\Lambda_n=n\beta \mu(\tau)$ and $\Theta=|\beta|^2 \lambda(\tau)$. Thus, after some rearrangements we obtain
\begin{equation}\label{19}
\begin{array}{rcl}
|\psi(\tau)\rangle=\sum\limits_{n,k}C_k D_n e^{in^2\Theta g(k)}|n\rangle_f\otimes|\Lambda_{n,k}(\tau)\rangle_m,
\end{array}
\end{equation}
where

\begin{eqnarray}\label{20}
|\Lambda_{n,k}(\tau)\rangle_m&=&f(k-1)!\sum\limits_{l,l'}\frac{(\Lambda_n)^l(-\Lambda_n^{*})^{l'}}{l!l'!}\times\\ \nonumber
&&\sqrt{\frac{(k+l)!}{(k-l')!}}
\frac{f(k+l-l'-1)!}{[f(k-l'-1)!]^2}|k+l-l'\rangle_m,
\end{eqnarray}
with $k\geq l'$. As is seen from Eqs. (\eqn{19}) and (\eqn{20}), if  the membrane is initially prepared in a vibrational
vacuum state, i.e., if $C_k=\delta_{k,0}$ then after a time $\tau=2\pi$ it returns to its original state. At all
times between  $\tau=0$  and $\tau=2\pi$ the membrane state is entangled with the cavity field
state. Furthermore, we see from equation (\eqn{19}) that there is an explicit Kerr-like term in
the time evolved state, so that physically one might expect  the cavity field to have an
evolution similar to that in a Kerr-like nonlinearity.

\section{NLCS of the movable membrane and its nonclassical properties}

\subsection{Generation of NLCS of the membrane}
In this subsection, we consider the generation of NLCS of a macroscopic membrane. For this purpose we turn our attention to the state (\eqn{19}). We assume a simple situation  $C_k=\delta_{k,0}$ that is the membrane is initially prepared in its motional ground state. Experimentally, the vacuum state of the membrane is feasible by cooling the membrane to its ground state\cite{thompson,jay1,jay}. Thus, the time dependent state (\eqn{19}) reduces to
\begin{equation}\label{21}
\begin{array}{rcl}
|\psi(\tau)\rangle=\sum\limits_{n} D_n e^{in^2\Theta g(0)}|n\rangle_f\otimes|\Lambda_{n}(\tau)\rangle_m,
\end{array}
\end{equation}
where $|\Lambda_{n}(\tau)\rangle_m\equiv |\Lambda_{n,0}(\tau)\rangle_m$. The explicit expression for the state $|\Lambda_{n}(\tau)\rangle_m$ in the number
representation is given by
\begin{eqnarray}\label{22}
|\Lambda_{n}(\tau)\rangle_m=e^{\Lambda_n \hat B^{\dagger}} |0\rangle_m
=\sum\limits_{l}|l\rangle_m {}_m\langle l|\Lambda_{n}(\tau)\rangle_m,
\end{eqnarray}
where
\begin{eqnarray}\label{23}
{}_m\left\langle {l}
 \mathrel{\left | {\vphantom {l {\Lambda _n (\tau )}}}
 \right. \kern-\nulldelimiterspace}
 {{\Lambda _n (\tau )}} \right\rangle _m=\aleph\frac{P(l)}{\sqrt{l!}}(\Lambda_n)^l,
\end{eqnarray}
and the normalization constant reads as
\begin{equation}
\begin{array}{rcl}
\aleph=\Big[\sum\limits_{l}\frac{|P(l)|^2}{l!}|\Lambda_n|^{2l}\Big]^{-\frac{1}{2}},
\end{array}
\end{equation}
with $P(l)=f(l-1)!=\prod\limits_{j=0}^{l-1}f(j),\,\,\,\,(l>0)$ and $P(0)=1$.\\
The density matrix of the system can be obtained by using Eq.(\eqn{21}),
\begin{eqnarray}\label{24}
\rho(\tau)&=&|\psi(\tau)\rangle\langle\psi(\tau)|\\
\nonumber &=&\sum\limits_{n,l} D_n D_l^{*}e^{i(n^2\Theta-l^2\Theta^{*}) g(0)}|n\rangle_f{}_f\langle l|\otimes|\Lambda_{n}(\tau)\rangle_m {}_m\langle \Lambda_{l}(\tau)|,
\end{eqnarray}
and the reduced density matrix of the motional state of the membrane reads as
\begin{eqnarray}\label{rhom}
\rho_m(\tau)&=&Tr_f[\rho(\tau)]\\
&=&\sum\limits_{n} |D_n|^2 e^{in^2(\Theta-\Theta^{*}) g(0)}|\Lambda_{n}(\tau)\rangle_m {}_m\langle \Lambda_{n}(\tau)|, \nonumber
\end{eqnarray}
where $|\Lambda_{n}(\tau)\rangle_m $ can be identified as a family of NLCSs of the membrane at time $\tau$ whose amplitude depends on $n$.
In order to get more clear insight to the above result we
present the following argument.

In principle, the nonlinear coherent states $|\zeta;f\rangle$ are defined as the right hand eigenstates of the generalized(deformed) annihilation operator $\hat B=f(\hat n) \hat b$\cite{dematos,manko,aniello},

\begin{eqnarray}\label{25}
\hat B|\zeta;f\rangle&=\zeta|\zeta;f\rangle,
\end{eqnarray}
where $f(\hat n)$ is a reasonably well-behaved real function of
the number operator $\hat n=\hat b^{\dagger}\hat b$ and $\zeta$ is an arbitrary complex
number. We may seek for an operator $\hat C^{\dagger}$ which is conjugate of the operator $\hat B$, that is
$[\hat B,\hat C^{\dagger}]=1$ while their Hermitian conjugates $\hat B^{\dagger}$ and  $\hat C$
satisfy the dual algebra $[\hat C,\hat B^{\dagger}]=1$. Thus it is easily
found that
\begin{eqnarray}\label{27}
\hat C=\frac{1}{f(\hat n)}\hat b ,\,\,\hat C^{\dagger}=\hat b^{\dagger}\frac{1}{f(\hat n)}.
\end{eqnarray}
Now, let us assume $D_n=\delta_{n,k}$ in Eq.(\eqn{21}), i.e., the cavity field is initially prepared in $|\psi(0)\rangle_f=|k\rangle_f$. Thus the reduced density matrix of the membrane reads as
\begin{eqnarray}\label{28}
\rho_m(\tau)&=&e^{ik^2(\Theta-\Theta^{*}) g(0)}|\Lambda_{k}(\tau)\rangle_m {}_m\langle \Lambda_{k}(\tau)|,
\end{eqnarray}
in which the state vector $|\Lambda_{k}(\tau)\rangle_m$  (given by Eq.(\eqn{22}))is the right-hand eigenstate of the deformed operator $\hat C$ given by Eq.(\eqn{27})
\begin{eqnarray}\label{29}
\hat C|\Lambda_{k}(\tau)\rangle_m&=\Lambda_{k}(\tau)|\Lambda_{k}(\tau)\rangle_m.
\end{eqnarray}
In order to investigate the nonclassical behaviour of the generated NLCS, we focus our attention on the temporal evolution of the
quadrature squeezing and of the Mandel parameter in the next subsection.
\subsection{Quadrature squeezing of the NLCS of the membrane}
Squeezed states are known by reduced quantum fluctuations in one quadrature of the field at the expense of the
increased fluctuations in the other quadrature. In order to investigate the quadrature squeezing of the generated nonlinear coherent state of Eq.(\eqn{22}) we define
\begin{eqnarray}\label{30}
\hat{X_1}(\tau)&\equiv &\frac{1}{2}(\hat{b} e^{i\tau}+\hat{b}^{\dag} e^{-i\tau}),\nonumber\\
\hat{X_2}(\tau)&\equiv &\frac{1}{2i}(\hat{b} e^{i \tau}-\hat{b}^{\dag} e^{-i \tau}).
\end{eqnarray}
These quadrature operators show the position and momentum operators of the mirror, respectively. They obey the commutation relation

\begin{equation}\label{31}
\begin{array}{rcl}
[\hat{X_1}(\tau),\hat{X_2}(\tau)]=i/2,
\end{array}
\end{equation}
and, consequently, the variances
 \begin{equation}\label{32}
\begin{array}{rcl}
<(\Delta\hat{X}_{j}(\tau))^2>\equiv<(\hat{X}_{j}(\tau))^2>-<(\hat{X}_{j}(\tau))>^2,\,\,(j=1,2)
\end{array}
\end{equation}
 satisfy the uncertainty relation
\begin{equation}\label{33}
\begin{array}{rcl}
 \left\langle {(\Delta \hat X_{1}^{} (\tau))^2 } \right\rangle \left\langle {(\Delta \hat X_{2}^{} (\tau))^2 } \right\rangle  \ge 1/16.
\end{array}
\end{equation}
A quantum state of the membrane is said to be squeezed when one of the quadrature components~$\hat{X}_{1}$ and~$\hat{X}_{2}$ satisfies the relation
\begin{equation}\label{34}
\begin{array}{rcl}
\Big<(\Delta\hat{X}_{j}(\tau))^2\Big> <1/4,\,\,(j=1\,or\,2).
\end{array}
\end{equation}
The degree of squeezing can be measured by the squeezing parameter $S_j(j=1,2)$ defined by
\begin{equation}\label{35}
\begin{array}{rcl}
S_j(\tau)\equiv 4<(\Delta\hat{X}_{j}(\tau))^2>-1.
\end{array}
\end{equation}
Then, the condition for squeezing in the quadrature component can be simply written as $S_j(\tau)<0$. The squeezing parameter $S_j(j=1,2)$ can be expressed in terms of the phonon annihilation and creation operators of the membrane as follows

\begin{eqnarray}\label{36}
S_1(\tau)&=&2A_1(\tau)+2Re[A_2(t)]-4(Re[A_3(\tau)])^2 ,\nonumber\\
S_2(\tau)&=&2A_1(\tau)-2Re[A_2(\tau)]-4(Im[A_3(\tau)])^2,
\end{eqnarray}

where
\begin{equation}\label{37}
\begin{array}{rcl}
A_1(\tau)\equiv<\hat{b}^\dag \hat{b}>,\, A_2(\tau)\equiv<\hat{b^2}>e^{2i \tau},\,\, A_3(\tau)\equiv<\hat{b}>e^{i \tau}.
\end{array}
\end{equation}

Now, we study the temporal behaviour of $S_2(\tau)$, which gives information on quadrature squeezing of
$X_2(\tau)$. Numerical results are presented in Fig.3, where we have
plotted $S_2(\tau)$ versus the scaled time $\tau$ for different values of the parameters $r_c$ and $\eta$. As is clear, the quadrature component $X_2(\tau)$ exhibits squeezing repeatedly at times $\tau=(2m+1)\pi$, $(m=0,1,2,...)$ and the amplitude of squeezing is strongly influenced by the Lamb-Dicke parameter $\eta$. Fig.3(a) shows that for $r_c=0.9$ the quadrature squeezing only exists for $\eta>0.19$ and the amplitude of squeezing increases by increasing the parameter $\eta$.
Furthermore, we find that with the increasing value of membrane's reflectivity, the quadrature squeezing is strengthened and it appears for smaller
values of $\eta$. For example, Fig.3(b) shows the quadrature squeezing of
$X_2(\tau)$ for $r_c=0.98$(this relatively high value of the membrane's reflectivity has recently been considered in Ref.\cite{kilic}). As is seen, for $\eta > 0.14$ the quadrature component $X_2(\tau)$ exhibits squeezing periodically in the course of time evolution.
\subsection{Mandel-Parameter}
The Mandel-Parameter corresponding to the NLCS of Eq.(\eqn{22}) is obtained as follows
 \begin{eqnarray}\label{233}
   M(\tau)&=&\frac{\langle \hat n_b^2(\tau)\rangle-\langle \hat n_b(\tau)\rangle^2}{\langle \hat n_b(\tau)\rangle}-1\\
  \nonumber &=&\frac{\sum\limits_{k}\frac{\aleph^2|\Lambda_{n}(\tau)|^{2k}k^2 P(k)^2}{k!}-\big(\sum\limits_{k}\frac{\aleph^2|\Lambda_{n}(\tau)|^{2k}k P(k)^2}{k!}\big)^2}
{\sum\limits_{k}\frac{\aleph^2|\Lambda_{n}(\tau)|^{2k}k P(k)^2}{k!}}-1.
\end{eqnarray}
This parameter vanishes for Poisson distribution, is positive for the super-Poisson distribution, and is negative for the sub-Poisson distribution.
Fig.4(a) shows the time evolution of the Mandel parameter with respect to the scaled time $\tau$ for two values of $\eta$ and for $r_c=0.9$. As is seen, the generated NLCS for the motional state of the membrane exhibits the sub-Poissonian statistics in the most of time. In addition, the sub-Poissonian characteristic is enhanced by increasing the Lamb-Dicke parameter. Fig.4(b) shows the behaviour of the Mandel parameter for a higher value of the membrane's reflectivity $r_c=0.98$. As is clear, the higher values of $r_c$ leads to a considerable enhancement of the sub-Poissonian characteristic.

\section{generation of superposition of the membrane's NLCS$\mathrm{s}$}
\subsection{Superposition of NLCSs with different amplitudes}
The system under consideration may also be used to generate a superposition of
separated NLCSs of the membrane. In order to verify this claim, we assume that the cavity field is initially prepared in a coherent state, i.e., $D_n=e^{-\frac{\mid\alpha\mid^2}{2}}\frac{\alpha^n}{\sqrt{n!}}$. Thus the reduced density matrix of the membrane, Eq.(\eqn{rhom}), takes the following form

\begin{eqnarray}\label{40}
\rho_m(\tau)&=&e^{-|\alpha|^2}\sum\limits_{n} \frac{|\alpha|^{2n}}{n!} e^{in^2(\Theta-\Theta^{*}) g(0)}|\Lambda_{n}(\tau)\rangle_m {}_m\langle \Lambda_{n}(\tau)|. \nonumber\\
&&
\end{eqnarray}
This state shows a superposition of NLCSs with different amplitudes $\Lambda_n$ for each value of $n$. The \textit{Q}-function of the motional state associated with the state(\eqn{40}), is given by

\begin{equation}
\begin{array}{rcl}
 Q(\tau)\equiv\frac{Q_m(\gamma_r,\gamma_i)}{\aleph^2}=\frac{1}{\pi\aleph^2}\langle\gamma|\rho_m(\tau)|\gamma\rangle
 =\frac{1}{\pi}e^{-|\alpha|^2-|\gamma|^2}
\sum\limits_{n}q_n,
\end{array}
\end{equation}
where
\begin{equation}\label{qn}
\begin{array}{rcl}
q_n=\frac{|\alpha|^{2n}}{n!} e^{in^2\Theta g(0)-in^2\Theta^* g(0)}\,\,\Big|1+\sum\limits_{l=1}\frac{(\Lambda_n|\gamma|e^{i\phi})^l}{l!}\Big|^2.
\end{array}
\end{equation}
Fig.5 illustrates the normalized \textit{Q}-function($Q(\tau)$) for different values of the Lamb-Dicke parameter and the membrane's reflectivity at time $\tau= 2.9$ and for $|\alpha|^2=4$. As is seen from these figures the separation between nonlinear coherent components is increased by increasing the parameters $\eta$ and $r_c$. Figures.5(a) and 5(b) show the double-peaked structure of the \textit{Q}-function. We see that how the separation between the two components can be controlled by varying the parameters $\eta$ and $r_c$. In other words, by varying $\eta$ and $r_c$ one can control the macroscopic distinguishability of the states involved in the superposition. Fig.5(c) shows an interesting case in which the membrane's state is a superposition of
two different kind of states with equal amplitude, one of them is a coherent like state and another is a squeezed like state. We mention that this state can only be prepared for large values of $\eta=0.98$ and $r_c=0.998$.
\subsection{Nonlinear Schr\"{o}dinger-like cat state}
In this subsection we consider the generation of the so-called nonlinear Schr\"{o}dinger cat state (NLSCS) of the membrane. For this purpose, we assume that the membrane is initially prepared in the NLCS discussed in section. III, i.e., $|\psi(0)\rangle_m=|\zeta;f\rangle_m=\sum\limits_{k}C_k|k\rangle_m$ with $C_k=\aleph'\frac{\zeta^k P(k)}{\sqrt{k!}}$, where the normalization constant is given by $\aleph'=\Big[\frac{|\zeta|^{2k} P(k)^2}{k!}\Big]^{-\frac{1}{2}}$. Thus the time dependent state of the system given by Eq.(\eqn{19}) at time $\tau=2\pi$ reads
\begin{eqnarray}\label{nlcs}
\big|\psi(\tau=2\pi)\big\rangle &=&\aleph'\sum\limits_{n,k}D_n e^{2\pi i n^2 |\beta|^2 g(k)}\frac{\zeta^k P(k)}{\sqrt{k!}}|n\rangle_f\otimes|k\rangle_m. \nonumber\\
&&
\end{eqnarray}
It should be noted that the experimental realization of the system under consideration shows $\eta<1$ and $\theta<<1$,(e,g. for the experimental values given in Refs.\cite{thompson,jay1,jay}, we obtain $10^{-2}\lesssim\eta<1$ and $\theta\lesssim10^{-3}$). Therefore, one may keep terms up to first order in the phonon number $\hat n_b$ in Eq.(\eqn{7}) and approximate the nonlinearity function $f_{(j=1)}(n_b)=f(n_b)$ by expanding the associated Laguerre polynomial
\begin{equation}\label{app}
\begin{array}{rcl}
f(\hat n_b)\simeq \epsilon+\sigma \hat n_b,
\end{array}
\end{equation}
where
\begin{widetext}
\begin{equation}
\begin{array}{rcl}
 \epsilon=\sum\limits_{m = 1} {\sum\limits_{k = 0}^{\frac{{m - 1}}{2}} {(m-2k)\frac{{\left| {r_c } \right|^m }}{m}} } \left( {\begin{array}{*{20}c}
   m  \\
   k  \\
\end{array}} \right)\frac{{(m - 1)!}}{{4^{m - 1} [(\frac{{m - 1}}{2})!]^2 }}\{ e^{-\frac{1}{2}(\eta\theta)^{2} (m - 2k)^{2}}\},\\
\\
\sigma=\sum\limits_{m = 1} {\sum\limits_{k = 0}^{\frac{{m - 1}}{2}} {\frac{(i \eta\theta)^2(m-2k)^3}{2!}\frac{{\left| {r_c } \right|^m }}{m}} } \left( {\begin{array}{*{20}c}
   m  \\
   k  \\
\end{array}} \right)\frac{{(m - 1)!}}{{4^{m - 1} [(\frac{{m - 1}}{2})!]^2 }}\{ e^{-\frac{1}{2}(\eta\theta)^{2} (m - 2k)^{2}}\}.
\end{array}
\end{equation}
\end{widetext}
Under this approximation, the nonlinear function $g(n_b)$ reduces to
 \begin{equation}\label{nonlinear}
\begin{array}{rcl}
g(k)\simeq\Gamma k^2-\Delta k+\epsilon^2,
\end{array}
\end{equation}
where we have defined
 \begin{equation}
\begin{array}{rcl}
\Gamma=(\epsilon-\sigma)^2+\sigma\epsilon+2\sigma;\,\,
\Delta=\sigma(\sigma-3\epsilon)+\epsilon^2.
\end{array}
\end{equation}
By using Eq.(\eqn{nonlinear}) in Eq.(\eqn{nlcs}), and assuming that the cavity field is initially prepared in the Fock state $|l\rangle$, $D_n=\delta_{n,l}$, the state of the membrane is obtained as
\begin{eqnarray}\label{nlcs1}
\big|\psi(2\pi)\big\rangle_m &=&\aleph'e^{2\pi i (l \epsilon|\beta|)^2}\sum\limits_{k} e^{2\pi i \xi k^2}\frac{(e^{i\varphi}\zeta)^k P(k)}{\sqrt{k!}}|k\rangle_m, \nonumber\\
&&
\end{eqnarray}
where
\begin{equation}
\begin{array}{rcl}
\phi=2\pi l^2 |\beta|^2\Delta,\,\,\xi=l^2 |\beta|^2\Gamma.
\end{array}
\end{equation}
Depending on the value of the parameter $\xi$, the state $|\psi(2\pi)\rangle_m $ can
be made equivalent to a variety of nonlinear multicomponent cat states. As an example for, $\xi=0.25$ we have
\begin{eqnarray}\label{nlcs2}
\big|\psi(2\pi)\big\rangle_m &=&e^{2\pi i (l \epsilon|\beta|)^2}\aleph_+\Big(e^{i\pi/4}|\zeta;f\rangle_m+e^{-i\pi/4}|-\zeta;f\rangle_m\Big).\nonumber\\
&&
\end{eqnarray}
where $\aleph_+=\big[2+2\aleph'^2\sum_{n=0}\frac{(-\zeta^2)^n P(n)^2}{n!}\big]^{-\frac{1}{2}}$ is the normalization constant\cite{mannn}.
 The state (\eqn{nlcs2}) shows a two-component NLSCS. The all higher order-multicomponent  nonlinear cats states of the motional state of the membrane may be obtained at time $\tau=2 \pi$ by varying the parameters $\xi$ and $\zeta$. The normalized $Q$-functions($Q(\tau)\equiv Q_m(\tau)/\aleph_{+}^2$) of the cat states produced by different values of $\xi$ and $\zeta$ and for a given value of the membrane's reflectivity are shown in Fig.6. It is interesting to note that by varying the membrane's reflectivity one can control the separation between the components of the generated NLSCS. To show this, we have plotted the normalized $Q$-function for three different values of $r_c$ in Fig.7. Obviously, by increasing the parameter $r_c$, the separation between the components of the produced NLSCS is increased.
\section{Effects of cavity field damping on the motional states of the membrane}
In order to make the model under consideration more realistic, we now take into account
the cavity field damping to examine its influence on the generated motional states of the
membrane. For this purpose, we consider a regime in which the radiation mode relaxes
much faster than the mirror. Experimentally, in the system under consideration the membrane Quality-factor is too large ($Q_m=10^6$\cite{thompson,jay1}) which leads to small membrane damping rate compare to the cavity damping $\kappa$. The case in which the mirror
relaxes much faster than the cavity mode, does not show
any quantum features due to the thermalization effects\cite{mancini3}. We assume that the number of thermal photons
is negligible at optical frequencies. Hence,
the master equation for the whole system reads as\cite{mancini3}
\begin{eqnarray}\label{lio1}
\hat{\dot{\rho}}(t)=\frac{i}{\hbar}[\hat{\rho},\hat{H}]+\textit{L}(\hat{\rho}),
\end{eqnarray}
with
\begin{eqnarray}\label{lio2}
\textit{L}(\hat{\rho})=\kappa/2(2\hat a \hat{\rho} \hat{a}^{\dagger}-\hat{a}^{\dagger}\hat{a}\hat\rho-\hat\rho \hat{a}^{\dagger}\hat a),
\end{eqnarray}
where $\kappa$ denotes the rate of damping due to the photon leakage from the cavity. In order to solve Eq.(\eqn{lio1}) we introduce the new density operator $\hat R$ as $\hat\rho(\tau)=\hat U(\tau)\hat R\hat U^{\dagger}(\tau)$. Here, $\hat U(\tau)$ is given by Eq.(\eqn{14}). By using this definition we rewrite the Eq.(\eqn{lio1})as

\begin{eqnarray}\label{lio5}
\hat{\dot{R}}=\hat{U}^{\dagger}\textit{L}(\hat U \hat R\hat{U}^{\dagger})\hat{U}^{\dagger}=\tilde{\textit{L}}(\hat R).
\end{eqnarray}
We now assume the above differential equation has a solution as $\hat R=\hat{R}_0+\hat R_1$, where $\hat R_0$ is a time independent operator, i.e.,$\hat {\dot{R_0}}=0$(this operator corresponds to the free solution of
Eq.(\eqn{lio1})). While, $\hat R_1$ satisfies the following time dependent differential equation
\begin{eqnarray}\label{lio}
\hat{\dot{R_1}}=\tilde{\textit{L}}(\hat R_0+\hat R_1).
\end{eqnarray}
Since the damping term $\textit{L}(\hat \rho)$ is small enough, one can apply the first order Born
approximation to replace $\tilde{\textit{L}}(\hat R_0+\hat R_1)$ by $\tilde{\textit{L}}(\hat R_0)$ in the right-hand side of Eq.(\eqn{lio}). Under
this approximation, the following solution for the operator $\hat R$ is obtained
\begin{eqnarray}
\hat R(\tau)=\hat R_0+\int_{0}^{\tau}\tilde{\textit{L}}(\hat R_0;t)dt.
\end{eqnarray}
Hence, the density matrix $\rho(t)$ is given by
\begin{eqnarray}\label{rhofi}
\hat \rho(\tau)=\hat \rho_0(\tau)+\hat \rho_1(\tau),
\end{eqnarray}
where
\begin{eqnarray}
\hat \rho_0(\tau)=\hat U(\tau)\hat\rho(0)\hat U^{\dagger}(\tau),
\end{eqnarray}
is the undamped part of the density matrix given by Eq.(\eqn{24}) and the damping part $\rho_1(\tau)$ has the following form
\begin{eqnarray}
\hat\rho_1(\tau)=\hat U(t)[\int_{0}^{\tau}\tilde{\textit{L}}(\hat R_0;t)dt] \hat U^{\dagger}(\tau).
\end{eqnarray}
To study the time evolution of the system in the presence of cavity damping we turn our attention to the density matrix of Eq.(\eqn{24}) in which the cavity field is assumed to be initially in a coherent superposition of Fock states, and
the membrane is assumed to be initially prepared in the motional ground state. By using
Eq.(\eqn{lio2}) the damping part $\hat \rho_1(\tau)$  reads as
\begin{eqnarray}\label{ro1}
\hat\rho_1(\tau)&=&\kappa\int_{0}^{\tau}[\tilde{\hat a}(\tau-t)\hat\rho_0(\tau) \tilde{\hat a}^{\dagger}(\tau-t)]dt\nonumber\\
&&-\kappa/2[\hat a^{\dagger}\hat a\hat\rho_0(\tau)+\hat\rho_0(\tau)\hat a^{\dagger}\hat a],
\end{eqnarray}
where we have defined
\begin{eqnarray}\label{aop}
\tilde{\hat a}(\tau-t)=\hat U(\tau-t)\hat a \hat U^{\dagger}(\tau-t).
\end{eqnarray}
Substituting from Eq.(\eqn{43}) into Eq.(\eqn{aop}) and applying the Baker-Campbell-Hausdorff formula together with using the Feynman disentangling theorem we arrive at

\begin{equation}\label{aop2}
\begin{array}{rcl}
\tilde{\hat a}(\tau-t)=\hat a e^{\beta \mu(\tau-t)\hat B^{\dagger}}e^{-\beta^{*} \mu^{*}(\tau-t) \hat B}e^{i |\beta|^2\lambda(\tau-t) g(\hat{n}_b)}.
\end{array}
\end{equation}
By substituting $\hat\rho_0(\tau)$  from Eq.(\eqn{24}) into Eqs.(\eqn{ro1}) we readily obtain
\begin{widetext}
\begin{eqnarray}
\hat\rho_1(\tau)&=&\kappa\sum\limits_{n,l} D_n D_l^{*}e^{i(n^2\Theta-l^2\Theta^{*}) g(0)}\int_{0}^{\tau}\Big[\big(\tilde{\hat a}(\tau-t)|n\rangle_f\otimes|\Lambda_{n}(\tau)\rangle_m \big)\big({}_f\langle l|\otimes{}_m\langle \Lambda_{l}(\tau)|\tilde{\hat a}^{\dagger}(\tau-t)\big)\Big]dt\nonumber\\
&&-\kappa/2\big[\sum\limits_{n,l}(n+l) D_n D_l^{*}e^{i(n^2\Theta-l^2\Theta^{*}) g(0)}|n\rangle_f{}_f\langle l|\otimes|\Lambda_{n,0}(\tau)\rangle_m {}_m\langle \Lambda_{l,0}(\tau)|\big].
\end{eqnarray}
\end{widetext}
Finally by using Eq.(\eqn{aop2}) we find the following expression for $\hat \rho_1(\tau)$

\begin{widetext}
\begin{eqnarray}\label{opfi}
\hat\rho_1(\tau)&=&\kappa\sum\limits_{n,l}\sum\limits_{k,k'} D_n D_l^{*}I_{n,l}^{k,k'}(\tau)\int_{0}^{\tau}\Big[e^{i|\beta|^2\lambda(\tau-t) g(k)}e^{-i|\beta|^2\lambda^{*}(\tau-t) g(k')}|n-1\rangle_f{}_f\langle l-1|\otimes|\Lambda_{1,k}(\tau-t)\rangle_m {}_m\langle \Lambda_{1,k'}(\tau-t)|\Big]dt\nonumber\\
&&-\kappa/2\big[\sum\limits_{n,l}(n+l) D_n D_l^{*}e^{i(n^2\Theta-l^2\Theta^{*}) g(0)}|n\rangle_f{}_f\langle l|\otimes|\Lambda_{n,0}(\tau)\rangle_m {}_m\langle \Lambda_{l,0}(\tau)|\big],
\end{eqnarray}
\end{widetext}
where by definition
\begin{equation}
\begin{array}{rcl}
I_{n,l}^{k,k'}(\tau)=e^{i(n^2\Theta-l^2\Theta^{*}) g(0)}\frac{\aleph^2\sqrt{n}\sqrt{l}(\Lambda_n(\tau))^{k}(\Lambda^{*}_l(\tau))^{k'}P(k)P(k')}{\sqrt{k!}\sqrt{k'!}},
\end{array}
\end{equation}
and $|\Lambda_{n,k}(\tau)\rangle_m $ is given by Eq.(\eqn{20}).

We are now in a position to examine the influence of the cavity field damping on the
motional state of the membrane by using the density matrix of the Eq.(\eqn{rhofi}). As before, we
assume the membrane to be prepared in a vibrational vacuum state initially. In Fig.8 we have
plotted the time evolution of the Mandel parameter of the membrane motional state for
various values of the rate of cavity field damping, the membrane's reflectivity and the Lamb-
Dicke parameter. From Fig.8(a) it is evident that unlike the undamped case ($\kappa=0$), at the
initial stages of evolution the Mandel parameter increases which shows super-Poissonian
statistics. However, as time goes on, this parameter decreases and the sub-Poissonian statistics
occurs. The rate with which the enhancement of super-Poissonian (sub-Poissonian) statistics
occurs is directly proportional to the membrane's reflectivity $r_c$; the greater $r_c$ is, the more
quickly the Mandel parameter increases (decreases). Furthermore, as in the undamped case,
with the increasing value of $r_c$ the sub-Poissonian characteristic of the motional state of the
membrane is strengthened. Also, Fig.8(b) shows that by increasing the lamb-Dicke parameter
$\eta$ the sub-Poissonian behaviour is enhanced. In Fig.9 we have illustrated the time evolution
of the squeezing parameter $S_2(\tau)$ in the presence of the cavity field damping. As is seen, the
photon leakage from the cavity suppresses the quadrature squeezing of the motional state of
the membrane in the course of time evolution. However, it is evident
that increasing the Lamb-Dicke parameter $\eta$ (Fig.9a) or the membrane's reflectivity $r_c$ (Fig.9b) brings about noise reduction in the quadrature $X_2$ and thus will increase the quadrature
squeezing. Thus, we conclude that the destructive effect of the cavity field damping on the
nonclassicality associated with the motional state of the membrane can effectively be
controlled by changing the parameters $r_c$ and $\eta$ .

As another aspect of the influence of the cavity field damping on motional state of the
membrane, we turn our attention to the NLSCS introduced in Sec.IV. For this purpose, we
consider the density matrix of Eq.(\eqn{rhofi}) at time $\tau=2\pi$ in which $\hat \rho_0(2\pi)$ corresponds to the
nonlinear multicomponent state of Eq.(\eqn{nlcs1}) and $\hat \rho_1(2\pi)$ is obtained from Eq.(\eqn{rhofi}). As an
example, the corresponding normalized quasiprobability \textit{Q}-function($Q_m(\tau)/\aleph^2$) at $\tau=2\pi$ for two different values of the
cavity field damping rate $\kappa$ and for given values of the parameters $r_c,\,\xi,\,\zeta$ is illustrated in Fig.10.
 As is seen, for small value of $\kappa$ the \textit{Q}-function consists of two well-separated peaks
(Fig.10(a)). By increasing the damping rate, one not only decreases the height of the two
peaks but also decreases the spatial separation between the NLCS components involved in the
superposition (Fig. 10(b)). In this connection, further numerical analysis reveals that by
changing the parameters $r_c$ and $\eta$ one can control the spatial separation between the
components of the generated NLSCS (i.e., controlling the influence of decoherence). It is also
worth noting that the generated motional NLSCS preserves more or less its quantum
coherence even in the presence of the cavity field damping. This survival of quantum coherence,
which can be attributed to the nonlinear character of the system under consideration, is in full
agreement with the fact that deformed (nonlinear) coherent states superposition can be more
robust against decoherence than the usual (nondeformed) Schr\"{o}dinger cat states \cite{mannn}. It has
been shown \cite{dematos} that deformed states, due to their nonlinear character, give rise to a more rich
phase space structure, part of which can easier survive against decoherence.

\section{summary and conclusions}
In this paper we have introduced a physical  scheme that allows one to generate and
control the nonclassical properties of motional nonlinear coherent states and their
superpositions for an undamped vibrating micromechanical membrane inside an
optical cavity. We have shown that if the cavity field is initially prepared in a Fock state
the motional state of the membrane may evolve to a family of nonlinear coherent states. We have been interested in analyzing the nonclassical properties of the generated state of
the membrane, including the quadrature squeezing and the sub-Poissonian statistics. In particular, we have found that the Lamb-Dicke parameter and the membrane's reflectivity lead to an enhancement of the nonclassical properties. As we have seen, with increasing the Lamb-Dicke parameter and the membrane's reflectivity the sub-Poissonian behaviour and quadrature squeezing of the motional state of the membrane are considerably strengthend. In addition, the scheme offers the possibility of generating various types of the so-called nonlinear multicomponent Schr\"{o}dinger cat sates of the membrane. We have shown that the separation between nonlinear coherent components is increased by increasing the parameters $\eta$ and $r_c$. We have also extended our treatment to a more realistic situation in which the photon leakage
from the cavity as a relevant source of decoherence is included and have examined its
influence on the nonclassical characteristics of the generated motional states of the membrane.
We have shown that it is possible to control the effect of the cavity field damping on the
nonclassical behaviour of the motional state of the membrane via the Lambe-Dicke parameter
and the membrane's reflectivity. In particular, we have found that the generated motional
NLSCSs of the membrane can be more robust against decoherence than the usual Schr\"{o}dinger
cat states.

At the end, we would like to point out
that our treatment is restricted  to the case of one-phonon excitations (j=1) of the
vibrational sideband. It is expected to generate some other interesting nonclassical
motional states of the membrane, e.g. nonlinear squeezed states, by considering higher
order of excitations. We hope to report on this issue in a forthcoming paper.
\section*{Acknowledgments}
The authors would like to express their gratitude to the
referees, whose valuable comments have improved the
paper. They are also grateful to the Office of Graduate
Studies of the University of Isfahan for their support.
\appendix
\section{Nonlinearity function $f_j(\hat n_b)$}
By applying the Baker-Campbell-Hausdorff theorem in Eq.(\eqn{3}) to disentangling the exponential terms and by using the series expansion of each exponential term we obtain
\begin{widetext}
\begin{equation}\label{a1}
\begin{array}{rcl}
\omega_c(\hat x)=\frac{\pi c}{2L}-\frac{c}{2L}
\sum\limits_{m = 1}^{\infty} {\sum\limits_{k = 0}^{\frac{{m - 1}}{2}} {\frac{{\left| {r_c } \right|^m }}{m}} } \left( {\begin{array}{*{20}c}
   m  \\
   k  \\
\end{array}} \right)\frac{{(m - 1)!}}{{4^{m - 1} [(\frac{{m - 1}}{2})!]^2 }}e^{-\frac{1}{2}(\eta\theta)^{2} (m - 2k)^{2}}\sum\limits_{l = 0}\{[(i\eta\theta) (m - 2k)]^{2l+j}\frac{(\hat b^{\dagger})^l\hat b^{l}}{l!(l+j)!}\hat b^j + h.c.\}.
\end{array}
\end{equation}
\end{widetext}
By making use of the relation
\begin{equation}\label{a2}
\begin{array}{rcl}
 (\hat b^{\dagger})^l\hat b^{l}=\frac{\hat n_b!}{(\hat n_b-l)!},
\end{array}
\end{equation}
Eq.(\eqn{a1}) can be written as

\begin{equation}\label{a3}
\begin{array}{rcl}
\omega_c(\hat x)=\frac{\pi c}{2L}-\frac{c}{2L}(i\eta\theta)^j\{f_j(\hat n_b)\hat b^j + h.c.\},
\end{array}
\end{equation}
where
\begin{widetext}
\begin{equation}\label{a4}
\begin{array}{rcl}
f_j(\hat n_b)=
\sum\limits_{m = 1} {\sum\limits_{k = 0}^{\frac{{m - 1}}{2}} {\frac{{\left| {r_c } \right|^m }}{m}} } \left( {\begin{array}{*{20}c}
   m  \\
   k  \\
\end{array}} \right)\frac{{(m - 1)!}}{{4^{m - 1} [(\frac{{m - 1}}{2})!]^2 }}\{ e^{-\frac{1}{2}(\eta\theta)^{2} (m - 2k)^{2}}\}\frac{\hat{n}_b!(m-2k)^{j}}{(\hat{n}_b+j)!}\Big[\sum\limits_{l = 0}\frac{[-(\eta\theta)^2 (m - 2k)^2]^{l}}{l!(l+j)!}\frac{(\hat n_b+j)!}{(\hat n_b-l)!}\Big].
\end{array}
\end{equation}
\end{widetext}
 Rewriting the above equation in terms of the associated Laguerre polynomials, $L_{n}^k(v)=\sum\limits_{m}\frac{(-v)^m(n+k)!}{(n-m)!(k+m)!m!}$, leads to Eq.(\eqn{7}).

\section{Disentangling of the time evolution operator by using Feynman operator calculus }
The unitary time evolution operator for the case of one-photon excitations $(j=1)$ can be written as
\begin{equation}\label{42}
\begin{array}{rcl}
\hat U^{(j=1)}(t)=e^{-i\omega_0\hat a^{\dagger}\hat a t}\hat U^{(j=1)}_{int}(t),
\end{array}
 \end{equation}
where
\begin{equation}\label{43}
\begin{array}{rcl}
\hat U^{(j=1)}_{int}(t)=\hat T exp\Big(-i\hat a^{\dagger}\hat a\int\limits_{0}^{t}(g\hat B^{\dagger}e^{i\omega_m s}+g^\ast\hat B e^{-i\omega_m s})ds\Big),
\end{array}
 \end{equation}
with $\hat B=f(\hat n)\hat b$ as the nonlinear(deformed) annihilation operator obeying the following commutation relation
\begin{eqnarray}\label{44}
[\hat B,\hat B^{\dagger}]=(\hat n+1)f^2(\hat n)-\hat nf^2(\hat n-1).
\end{eqnarray}
By using the Feynman operator calculus\cite{fey} we readily obtain
\begin{equation}\label{45}
\begin{array}{rcl}
\hat U^{(j=1)}_{int}(t)=e^{\beta\mu(t)\hat a^{\dagger}\hat a\hat B^{\dagger}}exp[-ig^\ast\hat a^{\dagger}\hat a \int\limits_{0}^{t}\hat B'(s)e^{-i\omega_m s}ds],
\end{array}
 \end{equation}
 where
\begin{equation}\label{46}
\begin{array}{rcl}
\hat B'(s)(s)=e^{-\beta\mu(s)\hat a^{\dagger}\hat a\hat B^{\dagger}}\,\hat B \,e^{\beta\mu(s)\hat a^{\dagger}\hat a\hat B^{\dagger}}.
\end{array}
 \end{equation}

By applying the Baker-Campbell-Hausdorff formula together with the commutation relation (\eqn{44}), Eq.(\eqn{46}) can be written as
\begin{equation}\label{47}
\begin{array}{rcl}
\hat B'(s)(s)=
\hat B + \sum\limits_{m = 1}^\infty  {\frac{{[- k(s)]^m }}{{m!}}} \hat {B^{\dagger}}^{m - 1} \sum\limits_{i = 0}^m {( - 1)^i } \frac{{m!}}{{i!(m - i)!}}F(\hat n + i),
\end{array}
 \end{equation}
where $k(s)=\beta\mu(s)\hat a^{\dagger}\hat a$ and $F(\hat n)=\hat n f^2(\hat n-1)$. Finally, by using the Feynman disentangling theorem we arrive at
\begin{equation}\label{47}
\begin{array}{rcl}
\hat {U}^{(j=1)}_{int}(\tau)= e^{\beta \mu(\tau)\hat a^{\dagger}\hat a \hat B^{\dagger}}e^{-\beta^{*} \mu^{*}(\tau)\hat a^{\dagger}\hat a \hat B}\times\\
\,\,\,\,e^{i |\beta|^2\lambda(\tau)(\hat a^{\dagger}\hat a)^2 g(\hat n_b)}e^{O(\beta ^3,\hat n_b)+...}.
\end{array}
\end{equation}

\bibliographystyle{apsrev}

\vskip 36cm
\section*{Figure Captions}
\textbf{Fig.1:} A high-finesse optical cavity with two rigid end mirrors and a dielectric membrane centered at the middle of cavity.\\
\\
\textbf{Fig.2:} The nonlinearity function $f(n_b)$ as a function of phonon number $n_b$ for: (a) $r_c=0.99$, $\eta=0.8$. (b)  $r_c=0.9$, $\eta=10^{-5}$. Here we have set $\theta= 10^{-4}$.\\
\\
\textbf{Fig.3:} Time evolution of the squeezing parameter $S_2(\tau)$, corresponding to the generated NLCS (\eqn{22}) versus the scaled time $\tau$: (a) for $r_c=0.9$ and $\eta=0.14$ (black line), $\eta=0.19$ (blue line), and $\eta=0.24$(green line).
  (b)for $r_c=0.98$ and $\eta=0.1$ (black line), $\eta=0.14$ (blue line), and $\eta=0.18$(green line). Here we have set $L=0.07$m, $m=50\,pg$.\\
\\
\textbf{Fig.4:} Time evolution of the Mandel parameter $M(\tau)$ corresponding to the NLCS (\eqn{22}) as a function of $\tau$:
(a) $r_c=0.9$, (b)~$r_c=0.98$ with~$\eta=0.25$(black line), and $\eta=0.3$(red line).\\
 \\
\textbf{Fig.5:} The normalized \textit{Q}-function($Q(\tau)\equiv Q_m(\tau)/\aleph^2$) of the generated state (\eqn{40}) of the membrane at time $\tau=2.9$: (a)~$r_c=0.95$,\,$\eta=0.8$; (b)~ $r_c=0.998$,\,$\eta=0.82$; (c)~$r_c=0.998$,\,$\eta=0.98$.\\
\\
\textbf{Fig.6:} The normalized $Q$-function($Q(\tau)\equiv Q_m(\tau)/\aleph_{+}^2$) of the multi-component cat states of the membrane at time $\tau=2\pi$ for different values of $\xi$ and $\zeta$, and for $r_c=0.95$: (a)$\xi=1.1,\,\,\zeta=0.25$; (b)$\xi=1.8,\,\,\zeta=0.25$; (c)$\xi=1.8,\,\,\zeta=\sqrt{\frac{1}{6}}$.\\
\\
\textbf{Fig.7:} The normalized $Q$-function($Q(\tau)\equiv Q_m(\tau)/\aleph_{+}^2$) of the multi-component cat states of the membrane at time $\tau=2\pi$ for different values
 of the membrane's reflectivity $r_c$, and for $\xi=1/\sqrt{8}$, $\zeta=1.8$: (a)$r_c=0.8.$ (b)$r_c=0.87.$ (c)$r_c=0.99.$\\
 \\
 \textbf{Fig.8:} Time evolution of the Mandel parameter $M(\tau)$ for different values of the cavity damping rate: (a)~$\eta=0.19$, and $r_c=0.93,\,\kappa=0$(black line), $r_c=0.93,\,\kappa=1\omega_m$(green line), $r_c=0.95,\,\kappa=1\omega_m$(blue line); (b)~$r_c=0.95$ and $\eta=0.19,\,\kappa=0$(black line), $\eta=0.19,\,\kappa=1\omega_m$(blue line), $\eta=0.16,\,\kappa=1\omega_m$(red line).
 \\
\\
\textbf{Fig.9:} Time evolution of the squeezing parameter $S_2(\tau)$ for different values of the cavity damping rate: (a)~$r_c=0.95$ and $\eta=0.19,\,\kappa=0$(black line), $\eta=0.19,\,\kappa=1\omega_m$(blue line), $\eta=0.16,\,\kappa=1\omega_m$(red line); (b)~$\eta=0.16$, and $r_c=0.96,\,\kappa=0$(black line), $r_c=0.96,\,\kappa=1\omega_m$(blue line), $r_c=0.95,\,\kappa=1\omega_m$(red line).
\\
\\
\textbf{Fig.10:} The normalized $Q$-function of the multi-component cat states of the membrane at time $\tau=2\pi$ in
the presence of cavity field damping for two different values of the damping rate $\kappa$, and for $r_c=0.95$, $\xi=1.8,\,\zeta=0.25$: (a)$\kappa=0.01\omega_m$, (b)$\kappa=0.4\omega_m.$
\\
\\
\\
\\
\\
\\

\vskip 46cm
\begin{figure}
    \begin{center}
    \includegraphics[width=2.7in]{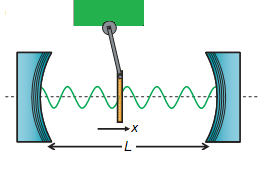}
   \caption{}
    \end{center}
\end{figure}
\begin{figure}
    \begin{center}
    \includegraphics[width=3.in]{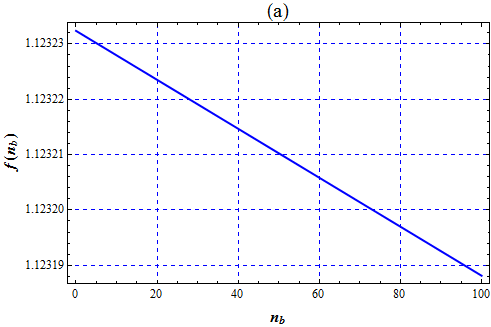}
        \vskip 2cm
      \includegraphics[width=3.in]{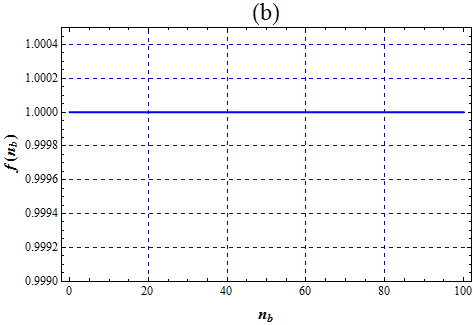}
   \caption{}
    \end{center}
\end{figure}
\begin{figure}
    \begin{center}
    \includegraphics[width=2.7in]{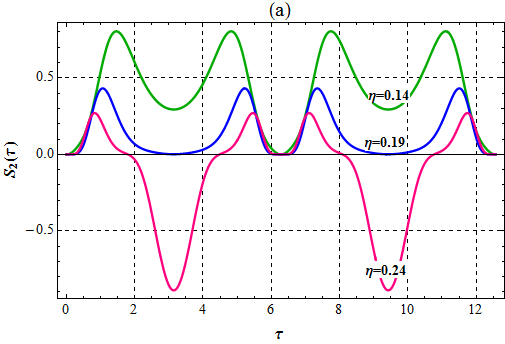}
        \vskip 2cm
      \includegraphics[width=2.7in]{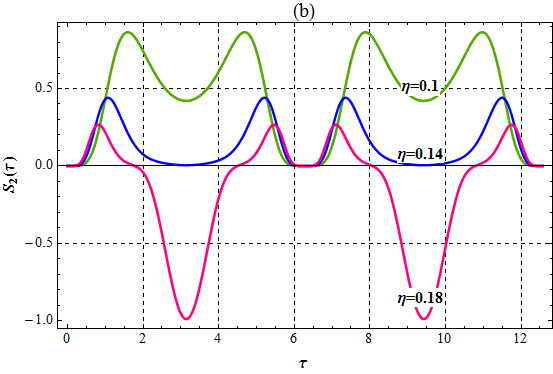}
   \caption{}
    \end{center}
\end{figure}
\begin{figure}
    \begin{center}
    \includegraphics[width=2.7in]{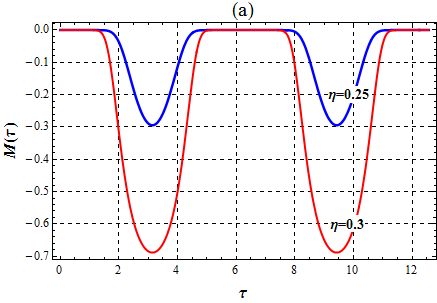}
    \vskip 2cm
      \includegraphics[width=2.7in]{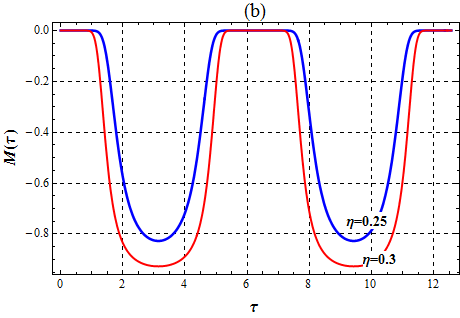}
       \caption{}
    \end{center}
    \end{figure}

\begin{figure}
    \begin{center}
    \includegraphics[width=3in]{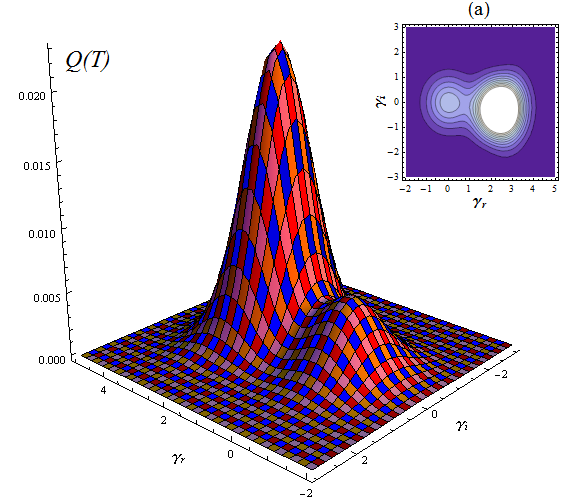}
    \vskip 0.5cm
      \includegraphics[width=3in]{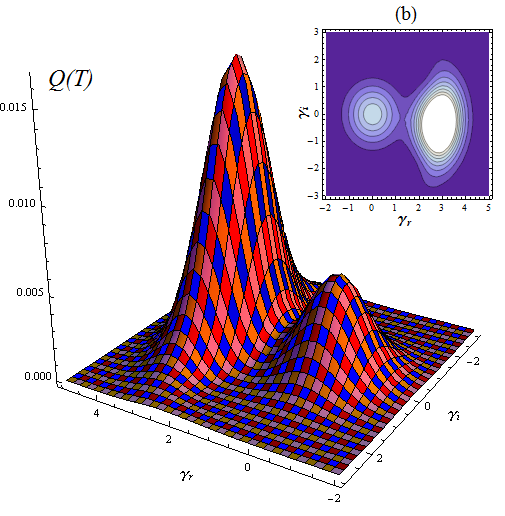}
          \vskip 0.5cm
  \includegraphics[width=3in]{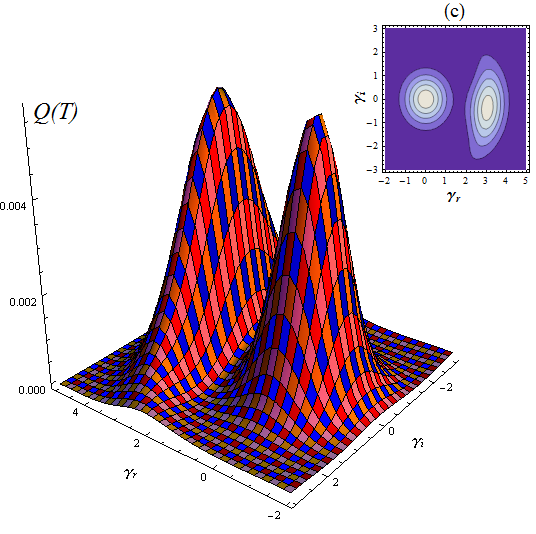}
       \caption{}
    \end{center}
\end{figure}
\begin{figure}
    \begin{center}
    \includegraphics[width=3.in]{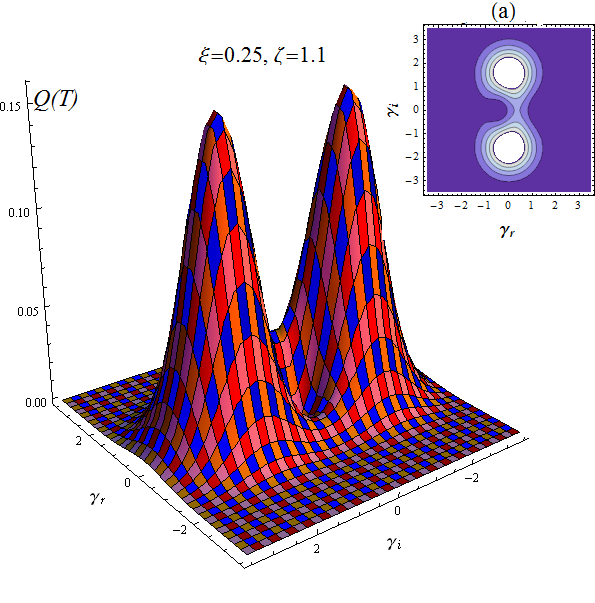}
    \vskip 1cm
      \includegraphics[width=3.in]{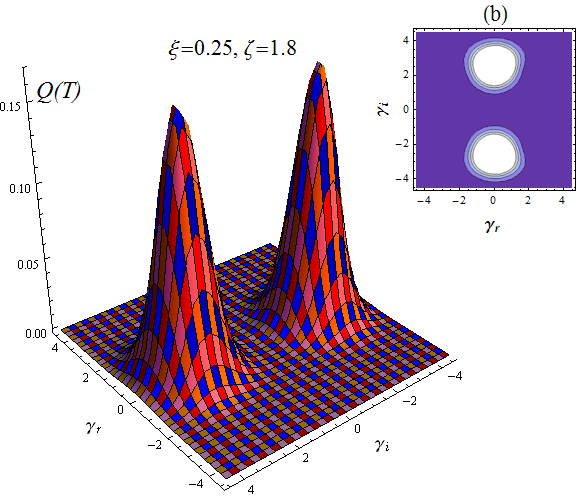}
          \vskip 1cm
  \includegraphics[width=3.in]{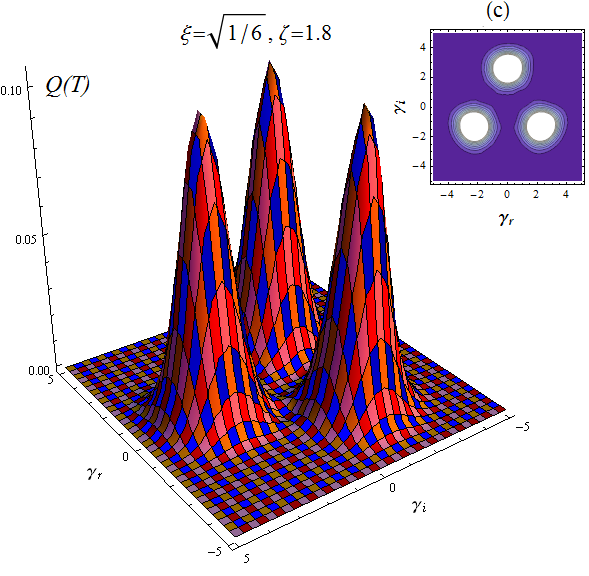}
       \caption{}
    \end{center}
\end{figure}
\begin{figure}
    \begin{center}
    \includegraphics[width=3in]{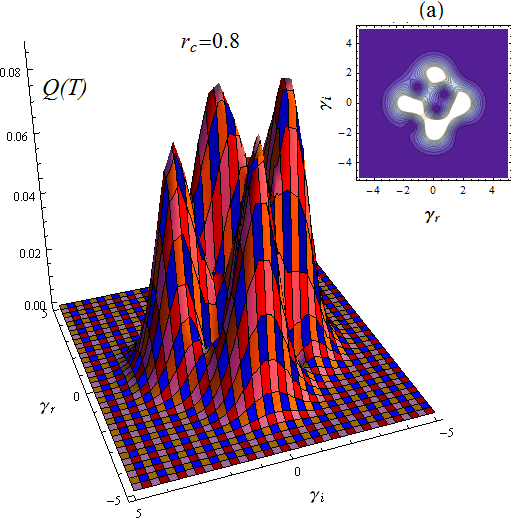}
    \vskip 1cm
      \includegraphics[width=3in]{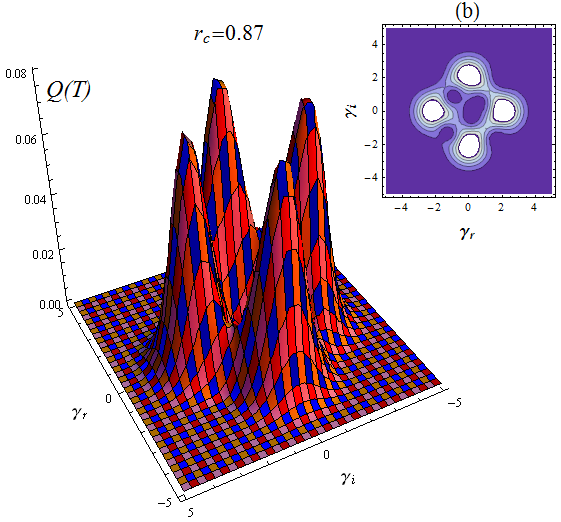}
          \vskip 1cm
  \includegraphics[width=3in]{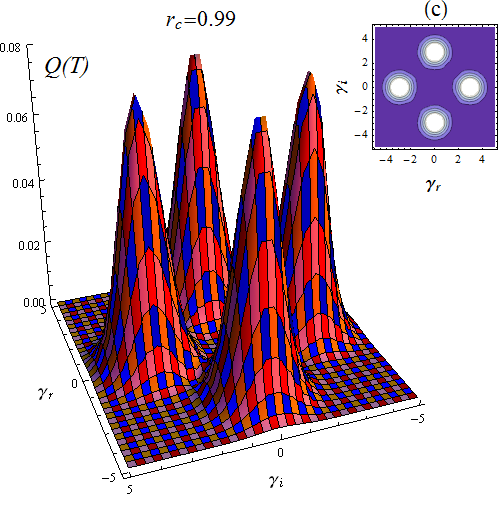}
       \caption{}
    \end{center}
\end{figure}
\begin{figure}
    \begin{center}
    \includegraphics[width=2.7in]{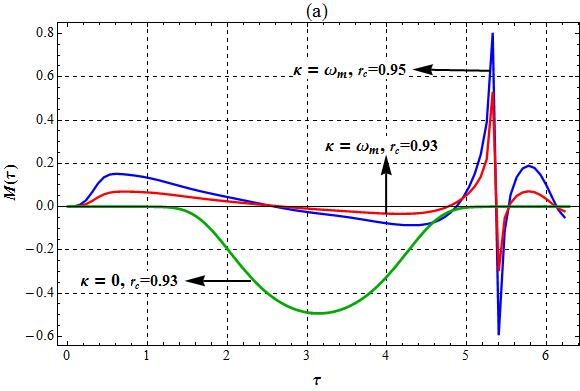}
    \vskip 1cm
    \includegraphics[width=2.7in]{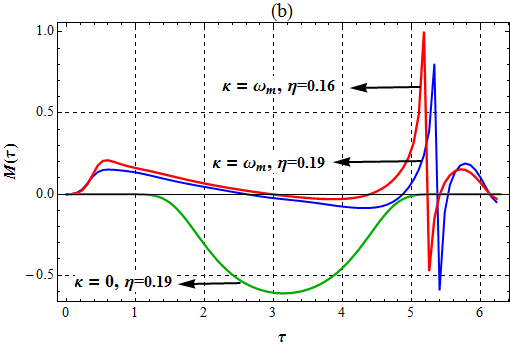}
       \caption{}
    \end{center}
\end{figure}
\begin{figure}
    \begin{center}
    \includegraphics[width=2.7in]{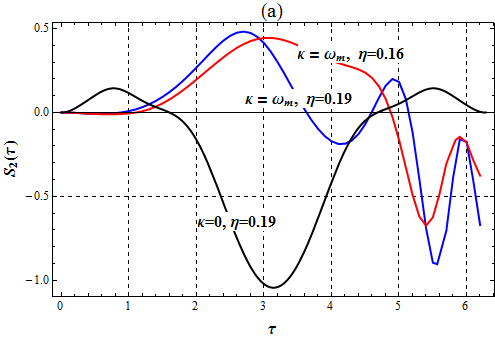}
    \vskip 1cm
    \includegraphics[width=2.7in]{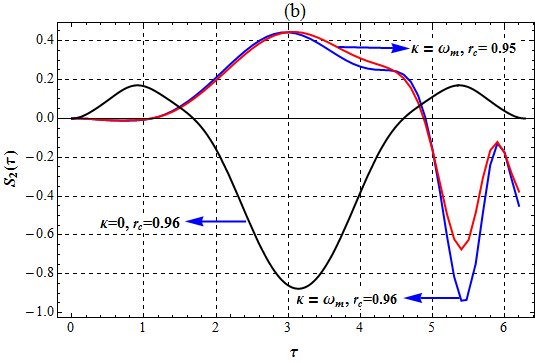}
       \caption{}
    \end{center}
\end{figure}
\begin{figure}
    \begin{center}
    \includegraphics[width=2.8in]{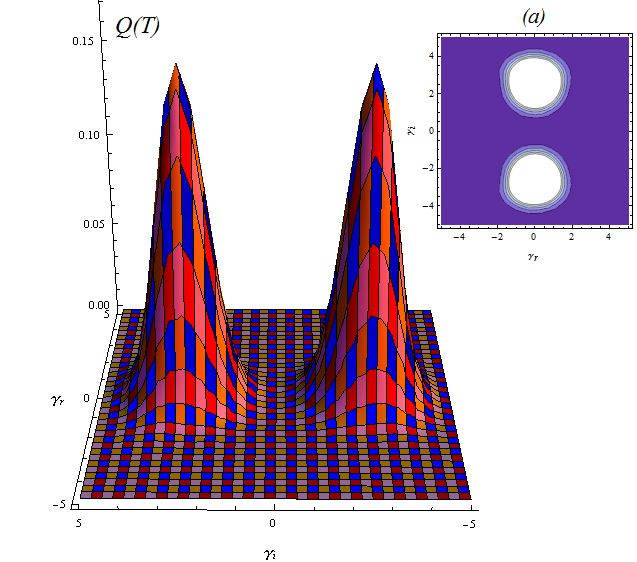}
    \vskip 1cm
      \includegraphics[width=2.8in]{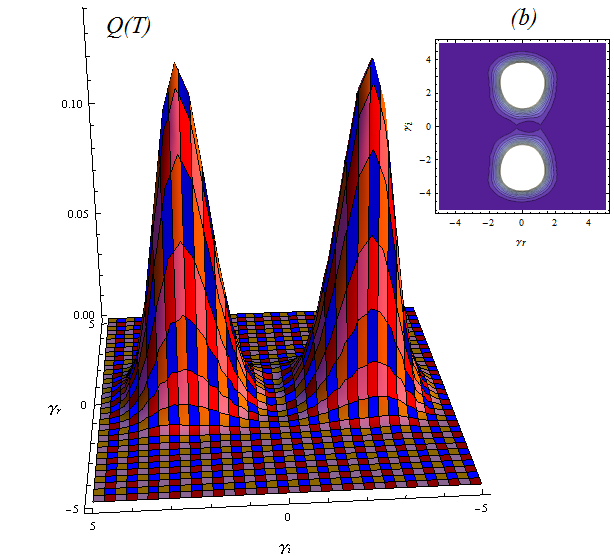}
       \caption{}
    \end{center}
\end{figure}
\end{document}